# MASS AND MACHIAN GENERAL RELATIVITY

Paul S. Wesson[1,2]

[1]Dept. Physics and Astronomy, University of Waterloo, Waterloo, Ontario N2L 3G1, Canada

[2]Space-Time-Matter Consortium, http://astro.uwaterloo.ca/~wesson

Correspondence: Mail to (1) above. Email= psw.papers@yahoo.ca

Abstract

Mach's Principle is usually taken to mean that the mass of a particle as measured locally is determined in some way by the other matter in the universe. This is difficult to formalize in 4D, but is feasible in 5D if the scalar potential of non-compactified Kaluza-Klein theory is interpreted as an inertial field. We therefore review 5D space-time-matter theory, but take the local particle mass to be defined by the integral of a global scalar field. This approach smoothly embeds general relativity, and leads to several new effects which can be tested.

1. Introduction

Inertia may be broadly defined as the resistance to acceleration of a test particle, and is measured by its rest mass. Mach suggested that this local property of a particle is somehow related to the other matter in the universe, an idea which motivated Einstein in his formulation of general relativity. But it is widely acknowledged that Mach's Principle is not fully incorporated into Einstein's field equations for four-dimensional gravitation. Neither does it play a meaningful role in the original Kaluza-Klein equations, which represent a restricted five-dimensional unification of gravitation and electromagnetism. However, in modern 5D theory, the fifth dimension is associated with a scalar field which can in principle be used to express particle mass. We think it is useful to re-express 5D space-time-matter theory, with the mass defined as an integral over the scalar field. This modest change of interpretation produces what is essentially a Machian version of general relativity.

One of the motivations for this is a recent account by Gogberashvili, in which he uses a Machian approach in 4D to resolve the hierarchy problem [1]. Basically, he argues that the countervailing influences of the many particles within the cosmic horizon causes the mass of a local particle to be reduced from the Planck value of order $10^{-5}$g to a more reasonable one. (For more conventional approaches to the hierarchy problem involving the Higgs field see ref. 2, and for approaches to Mach's Principle see ref. 3.) Several of the results from Gogberashvili's account are similar to ones found in the 5D approach to Kaluza-Klein field theory known as space-time-matter or STM theory [4]. There is a large literature on this and related theories, and we will use some of the key results below [5-20]. All 5D theories are motivated by the desire to extend 4D general relativity.

Einstein worked on 5D Kaluza-Klein theory in his later years. He was still motivated by Mach's suggestion of a link between microscopic and macroscopic physics; but also wished to rewrite his 4D field equations in a form where matter would be geometrized, turning the "base-wood" of the energy-momentum tensor into the "marble" of what we now call the Einstein tensor. Unfortunately, the form of the 5D field equations then in use was very restricted: the fifth or scalar potential was set to unity, all derivatives of the potentials with respect to the fifth coordinate were set to zero (the "cylinder" condition), and the topology of the fifth dimension was restricted to a circle ("compactification"). Given these hobbling conditions, it is hardly surprising that 5D relativity went little further than Kaluza's original "miracle", in which the 5D field equations gave back the 10 Einstein equations for gravitation and the 4 Maxwell equations for electromagnetism. Indeed, the 15th field equation – the one for the scalar field – was thrown away. With it went both the chance to geometrize matter, *and* (we will argue) the opportunity to formalize Mach's Principle.

So much for history. The modern approach to 5D relativity avoids the restrictions noted above, and is known generically as noncompactified Kaluza-Klein theory. It exists in two main versions, which are mathematically equivalent but differ in physical motivation [5]. Space-time-matter theory dates from 1992, and splits the 15 Ricci-flat field equations into sets of 10, 4 and 1 [6]. The first set is basically the Einstein equations of general relativity, but with an effective or induced energy-momentum tensor formed from all of the terms which depend on the fifth coordinate, thus realizing Einstein's dream of geometrizing matter. In other words, the fifth dimension is all around us, in the form of matter or energy. This interpretation is guaranteed to give back known 4D physics by virtue of Campbell's embedding theorem of differential geometry [7]. New physics comes from the other equations, which can be couched as a set of 4 conservation laws plus a wave equation for the scalar potential. It is the latter which will be the focus below. The other currently popular version of 5D relativity is membrane theory [8]. It dates from 1998, and is designed to explain the higher strengths of the particle-physics interactions compared to gravity, or equivalently the observed small masses of elementary particles compared to the theoretical Planck mass of order $10^{-5}$g (see above). It does this by constructing a 5D measure of distance (or interval) in which the 4D part is modulated by an exponential function, involving the extra coordinate and a length scale related to the cosmological constant. This causes the interactions of particles to be concentrated on a singular hypersurface (the "brane"), whereas gravity propagates outside and is so diluted in the external region (the "bulk"). In other words, the fifth dimension exists, but most of laboratory physics happens on the 4D hypersurface we call spacetime.

In what follows, we will not be much concerned with the contending philosophies underlying the two main approaches to modern 5D relativity, because our purpose is to utilize their common mathematical formalism to investigate the physics of Mach's Principle. Specifically, our goal is to realize this Principle by giving an expression for the (inertial) rest mass of a local test particle in terms of a global scalar potential. However, that potential must be determined by solving a set of field equations, and this in turn depends on a choice of coordinate frame, or gauge. In this regard, we have found that it is more useful to employ the canonical metric of space-time-matter theory than the warp metric of membrane theory. However, we do take from the latter approach the premise that 5D relativity is a classical field theory whose quantum counterpart involves a spin-2 graviton, a spin-1 photon and a spin-0 scaleron. It is the last which we believe to be instrumental in fixing particle masses. In a quantum-field approach to this problem, masses are fixed by the Higgs mechanism [2, 9]; but in our account we cannot address this issue in detail, because at the classical level the fields are all of infinite range and therefore involve quanta which are technically massless. Even so, we will venture some comments about testing our approach using the Large Hadron Collider, which are based on classical relations. In fact, much of what we will do is based on results from the classical theory which are already in the literature [10-20]. Our purpose is to reinterpret these, to provide a consistent Machian theory of relativity, and to list new effects which can be used to test this.

2. Mass as a Field

That new physics can come from an extra dimension is apparent, if for no other reason than that to an observer in 4D the extra coordinate and functions of it may correspond to measurable effects. An example is the so-called fifth force, which depends on the relative velocity between the 4D and 5D frames, and causes an inertial acceleration analogous to that felt by a person on a roundabout which moves with respect to the Earth. It has been isolated in both space-time-matter theory and membrane theory [10, 11]. In both approaches, it is also known that massive particles moving on timelike paths in 4D may be moving on null paths in 5D, a result we will use below. In general, deriving new physics in 4D from the algebra of 5D involves choosing coordinates, writing down the equations of the theory, and solving these with an appropriate physical interpretation.

Coordinates have to be chosen relevant to the physics in view. This is not trivial, especially in 5D. Thus membrane theory commonly uses the warp metric to concentrate particle interactions in its singular hypersurface [8, 9, 11]. While STM theory commonly uses the canonical metric, which preserves momentum under the action of the fifth force noted above [10, 12, 16]. We wish to retain the conventional definition of momentum, as the product of (inertial) rest mass m and the 4-velocity $u^\alpha \equiv dx^\alpha / ds$, defined using the regular 4D coordinates $x^\alpha$ and the 4D proper time $s$. In fact, we wish to make contact with the large body of standard physics which is expressed in terms of the 4D proper time, so we will use $s$ rather than its 5D counterpart $S$ as dynamical parameter. Our goal is to obtain a meaningful theory for the rest mass $m$, instead of merely accepting it as a given scalar quantity like in 4D general relativity. Indeed, the problem of Mach's Principle could be said to be solved if we can obtain from 5D field theory an expression of the form $m = m(x^\alpha, x^4)$. Here $x^\alpha$ refers to the 4D coordinates of time ($x^0$) and space ($x^{123}$), and $x^4$ is a measure for the extra dimension. Henceforth, we will label the 5D coordinates $x^A = (x^\alpha, \ell)$, where $x^4 = \ell$ is chosen to avoid confusion with the usual Cartesian length, and to discourage the implication that the extra coordinate is measured with respect to any special hypersurface (membrane theory commonly uses $x^4 = y$ with the brane at $y = 0$). The new axis $x^4$ is locally orthogonal to the 4D surface in $x^\alpha$ which defines spacetime. The latter has line element $ds^2 = g_{\alpha\beta} dx^\alpha dx^\beta$ defined as usual with a 4D metric tensor, and is contained in the corresponding 5D line element $dS^2 = g_{AB} dx^A dx^B$ ($A, B = 0, 123, 4$). The 5D metric tensor consists of the 4D one plus extra terms, where the scalar field is described by the extra diagonal component $g_{44}$. We are not here concerned with electromagnetism, which is traditionally described by the 4 off-diagonal components $g_{\alpha 4}$ [4]. Therefore, we use 4 of the available 5 degrees of coordinate freedom in the metric to set $g_{\alpha 4} = 0$ and remove the Maxwell potentials (which are frequently defined as $A_\alpha \equiv g_{\alpha 4} / g_{44}$). For algebraic convenience, we also write $g_{44} = \varepsilon \Phi^2$, where $\varepsilon = \pm 1$ allows for both a spacelike and timelike extra dimension, and $\Phi = \Phi(x^\alpha, \ell)$ is the scalar potential with which we are mainly concerned. In sum, we have a 5D interval given by

$$dS^2 = g_{\alpha\beta}\left(x^\gamma, \ell\right) dx^\alpha dx^\beta + \varepsilon \Phi^2\left(x^\gamma, \ell\right) \quad . \tag{1}$$

This is general. It can be specialized to the canonical form to describe the dynamics of variable-mass particles (see below), or to the warp form to describe the confinement of massive particles to a singular hypersurface.

Field equations for a metric like (1) will in general number 15. For while it has only 11 nonzero potentials (the 10 components of the symmetric tensor $g_{\alpha\beta}$ and the 1 component of the scalar field $\Phi$), these contribute to all components of a 5D geometrical quantity such as the Ricci tensor $R_{AB}$. The latter is in fact the primary object in modern noncompactified Kaluza-Klein theory. It is the 5D analog of the 4D Ricci tensor $R_{\alpha\beta}$, which with its associated scalar $R$ is used to construct the Einstein tensor $G_{\alpha\beta} \equiv R_{\alpha\beta} - (R/2) g_{\alpha\beta}$. This of course forms the left-hand-side of Einstein's field equations of general relativity, whose right-hand-side contains the properties of matter in the form of the energy-momentum tensor $T_{\alpha\beta}$. The full form of Einstein's equations is $G_{\alpha\beta} = 8\pi T_{\alpha\beta}$. (Here units are chosen so that the gravitational constant $G$ and the speed of light $c$ are unity.) However, Einstein's equations in the absence of matter reduce to $R_{\alpha\beta} = 0$, which is the form verified by the classical tests of relativity. More importantly, it is now known that Einstein's equations with matter in 4D are a subset of the Ricci-flat equations for apparently empty space in 5D, where the source terms in the former correspond to the extra terms in the latter. This is a consequence of Campbell's embedding theorem, and is the reason why space-time-matter theory is sometimes called induced-matter theory [7]. Irrespective of this, many workers have anyway chosen to adopt the simplest field equations in 5D, which are the Ricci-flat ones. These read

$$R_{AB} = 0 \quad (A, B = 0, 123, 4) \quad . \tag{2}$$

The 15 components of this may be expanded, and grouped into sets of 10, 4 and 1 as noted previously.

The first set of (2) is Einstein-like, and it is instructive to separate terms: the standard, purely 4D ones *versus* the terms which involve the new potential $\Phi$ or derivatives of the other potentials with respect to the new coordinate $\ell$. This enables us to obtain expressions for standard 4D quantities in terms of those which define the 5D embedding. An example is the 4D Ricci scalar, which in terms of the extra dimension is given by

$$R = \frac{\varepsilon}{4\Phi^2} \left[ g^{\mu\nu}{}_{,4} g_{\mu\nu,4} + \left( g^{\mu\nu} g_{\mu\nu,4} \right)^2 \right] \quad . \tag{3}$$

Here a comma denotes the partial derivative, and below we will use a semicolon to denote the regular 4D covariant derivative. Using (3) and similar relations, the first 10 components of the 5D field equations (2) take the form of Einstein's relations:

$$G_{\alpha\beta} = 8\pi T_{\alpha\beta}$$

$$8\pi T_{\alpha\beta} \equiv \frac{\Phi_{,\alpha;\beta}}{\Phi} - \frac{\varepsilon}{2\Phi^2} \left\{ \frac{\Phi_{,4} g_{\alpha\beta,4}}{\Phi} - g_{\alpha\beta,44} + g^{\lambda\mu} g_{\alpha\lambda,4} g_{\beta\mu,4} \right.$$

$$\left. - \frac{g^{\mu\nu} g_{\mu\nu,4} g_{\alpha\beta,4}}{2} + \frac{g_{\alpha\beta}}{4} \left[ g^{\mu\nu}{}_{,4} g_{\mu\nu,4} + \left( g^{\mu\nu} g_{\mu\nu,4} \right)^2 \right] \right\} \quad . \tag{4}$$

The effective energy-momentum tensor here contains contributions from the scalar potential ($\Phi$) and contributions from derivatives with respect to $\ell$ of the 4D potentials ($g_{\alpha\beta}$). It is known that (4) gives back all known forms of ordinary matter, which again can be viewed as a consequence of Campbell's theorem [6, 7, 13, 14]. In this manner, the 10 relations in (4) automatically satisfy the majority of the 5D field equations.

The remaining 5 components of (2) have expressly to do with 5D, and have no 4D counterparts. The $R_{\alpha 4} = 0$ components of (2) can be most conveniently expressed in terms of 4 conservation relations for a new tensor, thus:

$$P^{\beta}_{\alpha;\beta} = 0$$

$$P^{\beta}_{\alpha} \equiv \frac{1}{2\Phi}\left(g^{\beta\sigma}g_{\sigma\alpha,4} - \delta^{\beta}_{\alpha}g^{\mu\nu}g_{\mu\nu,4}\right) \quad . \tag{5}$$

This tensor has an associated scalar $P = -3g^{\lambda\sigma}g_{\lambda\sigma,4}/2\Phi$. It is possible to relate this to the rest mass of a particle in gauges of canonical type (see below), so the 4 relations above can be viewed as conservation laws for a kind of mass current. They are, in effect, the neutral analogs of the conservation laws for the electric current in the more general case where the metric (1) has electromagnetic potentials. In the neutral case we are studying here, the covariant form $P_{\alpha\beta}$ of the mixed tensor in (5) is by some workers added to the regular energy-momentum tensor $T_{\alpha\beta}$, thereby forming a composite source for the hypersurface in membrane theory [4, 5, 9, 13]. In many situations, the relations (5) are quite easy to satisfy.

The same comment applies to the last component of (2), namely $R_{44} = 0$. Written out in full, this reads

$$\Box\Phi = -\frac{\varepsilon}{2\Phi}\left[\frac{g^{\lambda\beta}_{,4}g_{\lambda\beta,4}}{2} + g^{\lambda\beta}g_{\lambda\beta,44} - \frac{\Phi_{,4}g^{\lambda\beta}g_{\lambda\beta,4}}{\Phi}\right]$$

$$\Box\Phi \equiv g^{\alpha\beta}\Phi_{,\alpha;\beta} \quad . \tag{6}$$

This is perhaps the simplest of the 5D field equations to interpret: it is a wave equation for the scalar field $g_{44} = \varepsilon \Phi^2(x^\gamma, \ell)$ of metric (1). It is sourceless if the 4D metric tensor is independent of $x^4 = \ell$, but sourceful otherwise. Thus the scalar field of 5D Kaluza-Klein theory is similar to the gravitational field of Newtonian theory, being determined by the equations of Laplace or Poisson, depending on the absence or presence of matter. We will see below that (6) is central to a Machian theory of mass.

Many solutions are known of the field equations (4)-(6). Some like the isolated objects known as solitons do not depend on $x^4 = \ell$ [15, 16], while others like the standard 5D cosmologies do (see ref. 14 for a compendium of solutions). The field equations agree with all extant observational data for the solar system and other large-scale astrophysical systems [4]. The same applies to the laws of motion which follow from metric (1), either by extremizing the interval or by setting it to zero to describe 5D null paths [10, 11]. The theory is compatible with the Weak Equivalence Principle, which shows the proportionality of inertial (i.e. dynamical) and gravitational mass to 1 part in $10^{12}$ [12, 13]. We can use the WEP to justify the use of a single scalar quantity $m$ to describe the rather slippery concept of mass, though our aim is to considerably widen our understanding of it.

To do this, we rewrite metric (1) in a form which brings out the new coordinate in a manner analogous to what is done in standard 4D cosmology. There, the 4D line element is written $ds^2 = dt^2 - R^2(t)d\sigma^2$, where $t$ is a synchronous time whose value is agreed on by all observers, and $R(t)$ is a scale factor which determines the evolution of a curved 3D subspace $d\sigma^2$ in which the spatial coordinates are comoving with the matter and are in the nature of labels. The 'real' or proper distance in 3D is $\int R(t)d\sigma$, which is particularly simple for the Milne model where $R(t) = t$ (we are absorbing constants). Transferring these considerations from 4D to 5D, and choosing a spacelike extra dimension, we can rewrite (1) in the form

$$dS^2 = (\ell^2/L^2)g_{\alpha\beta}(x^\gamma, \ell)dx^\alpha dx^\beta - \Phi^2(x^\gamma, \ell)d\ell^2 \qquad (7)$$

$$= (\ell^2/L^2)ds^2 - \Phi^2 d\ell^2 \qquad . \qquad (8)$$

This is algebraically still general, and indeed has one degree of coordinate freedom in reserve. This is sometimes used to suppress the scalar potential by the choice $g_{44} = -\Phi^2 = -1$. In this gauge with $g_{\alpha\beta} = g_{\alpha\beta}(x^\gamma$ only), the constant length $L$ introduced to (7) for the consistency of physical dimensions can be identified from the Einstein-like field equations (4) in terms of the cosmological constant of general relativity, via $\Lambda = 3/L^2$ [16]. In this gauge, the field equations and the equations of motion are much simplified, so (7) with $\Phi = 1$ is sometimes called the canonical gauge. It has a considerable literature [4, 14, 16]. We will below consider the general case of $\Phi = \Phi(x^\gamma, \ell)$; but here we wish to go to the literature on metrics like (7), and extract some results which bear on the interpretation of the extra coordinate $x^4 = \ell$ in a physical sense.

There are five known reasons for believing that $\ell$ is related to the rest mass $m$ of a test particle: (a) If $\ell$ is proportional to $m$ in the metric (7), its first part gives back the element of action for conventional 4D dynamics, namely $mds$. (b) The laws of motion which follow from the 5D interval are consistent with the conservation of the conventionally-defined 4D momentum, if $\ell$ is proportional to $m$. (c) The conserved quantity associated with the time axis of the metric (7) is, in the appropriate limit, $\ell(1-v^2)^{-1/2}$ where $v$ is the 3-velocity, which is the conventional 4D quantity if $\ell = m$. (d) The effective 4D energy-momentum tensor (4) which follows from the 5D field equations (2) is traceless when the 5D metric does not depend on $\ell$ (even via a quadratic factor), so the equation of state for massless particles follows from the absence of $\ell$ in the metric, as expected if $\ell$ measures $m$. (e) Conventional mechanics is based on three independent physical dimensions (*M, L, T*), so we might on general principles expect mass to require its own coordinate, if it is variable like space and time. The preceding five points may not constitute a proof that $x^4 = \ell = m$; but together they form a reasonable case for believing that this presents a viable way of implementing Mach's Principle.

We therefore proceed on the assumption that (7) represents a momentum manifold, rather than just an extension of a coordinate manifold. Recalling the similarities between 4D and 5D metrics drawn above, it is apparent that the general definition of rest mass involves the analog of the proper distance in the fifth dimension. We accordingly define

$$m \equiv \int \Phi(x^\gamma, \ell) d\ell \quad . \tag{9}$$

This for its evaluation requires a solution of the field equations (2), with the metric in the form (7). There are several such solutions in the literature, which give back expressions for (9) with reasonable physical properties [4, 14]. Several consequences of (9) were studied by Ma [17], who realized its Machian character and found agreement with observations in a number of applications. However, it should be noted that a coordinate transformation away from the quadratic form (7) will not in general leave intact the momentum-conserving nature of the metric which we have been concerned to establish. This because the group of 4D coordinate transformations $x^\alpha \to \bar{x}^\alpha(x^\beta)$ is narrower than the 5D group $x^A \to \bar{x}^A(x^B)$, so a 4D-defined quantity like the momentum will in general alter its form under a change of 5D gauge. It should also be noted that practical physics often involves the mass of a test particle as measured using the 4D proper time *s* along some specific path, determined now by the 5D equations of motion. The latter are well known for metrics like (7) which lack electromagnetic potentials (see e.g. ref. 10). They are also known for more general metrics where there are Maxwell-like potentials, which can be re-introduced to (7) via an $\ell$-dependent coordinate transformation. (See e.g. ref. 14 p.165 and ref. 19; particles will in general move away from a nonsingular hypersurface unless constrained by a nongravitational force such as electromagnetism, but the rate is small, being governed for canonical-type metrics by the cosmological constant.) In the case where the rest mass of a test particle is determined in a dynamical manner along its path, an alternate form of (8) is

$$m \equiv \int \Phi(x^\gamma, \ell) w \, ds \quad , \tag{10}$$

where $w \equiv d\ell/ds$ is the particle's velocity in the extra dimension. Whether (9) or (10) is used, it is clear that the rest mass of a test particle as defined above depends on the scalar function $\Phi$, which is therefore a kind of mass field. Locally, or in the weak-field limit, we have $\Phi \to 1$. This brings us back to the canonical metric discussed above, where $m = \ell$ is already shown by several results in the literature.

In other words, Mach's Principle is realized by defining the mass of a local object as the integral of a global scalar field over the extra dimension. This approach, defined by (9) or (10), gives the ordinary mass $m$ when the field $\Phi$ is a constant in spacetime ($x^\alpha$) and the extra coordinate ($x^4 = \ell$). It works as a first-order description of mass; and a preliminary analysis of more significant forms for the scalar field shows that in general the modifications involve the ratio of the test mass to the source mass [4, 12]. However, it is clear that to properly examine the influence of the mass field, we need to look at more extreme situations.

This enables us to make some predictions:

(i) The soliton solutions of the field equations (2) represent isolated, 3D spherically-symmetric systems which for want of a better word we call particles [15, 16]. Most of them have significant scalar fields, which by the $\Phi$-terms in the effective energy-momentum tensor (4) are uniquely 5D in origin. For the simplest such case, there are terms in $T_1^1$ and $T_2^2 = T_3^3$ which fall off like $1/r^3$ with radius [20]. These may represent stabilizing pressures for conventional particles or stresses for exotic objects like black strings, but in any case are in principle measurable.

(ii) The cosmological constant also measures the energy density of the vacuum fields near particles [4]. When a metric has the form (7) and the scalar field is approximately constant, we noted above that $\Lambda$ as determined by the Einstein field equations (4) is given in terms of the length scale of the spacetime by $\Lambda = 3/L^2$ [16]. However, if we shift the extra mass-related coordinate by a constant so $\ell \to (\ell - \ell_0)$, we obtain $\Lambda = (3/L^2)\ell^2(\ell-\ell_0)^{-2}$ [18]. Applied to particle physics, this describes a resonance which depends on the interaction energy. Applied to cosmology, where in general $\ell$ varies with proper time, it implies that the cosmological 'constant' could have diverged when $\ell = \ell_0$, or at the big bang.

(iii) For 5D cosmological solutions of the field equations (2) with the Machian interpretation of the scalar field we have outlined above, particle masses will in general vary slowly with cosmic time [4, 14]. Other things being equal, this will affect quantities like the Thomson cross-section for the electron $(8\pi/3)(e^2/mc^2)^2$, which determines the thermalization history of cosmic radiation. Observations at high redshift can be used to test the underlying theory in these ways.

(iv) For both local and cosmological physics, the relation (3) measures the 4D curvature $R$ in terms of the 5D embedding. This is complementary to the traditional measure, which uses quantities confined to the 4D hypersurface we call spacetime. This brings up an interesting general issue for discussion: how do we characterize the results of experiments done in 4D when they sample 5D? In the case of (3), it gives the 4D curvature in terms of derivatives of the potentials along the extra axis (orthogonal to spacetime) and the potential associated with the extra dimension. We suggest that a convenient way to characterize the curvature measure (3) is by a broad redefinition of the cosmological 'constant' $\Lambda$, using the 4D vacuum Einstein relation $|R| = 4|\Lambda|$. This essentially uses the energy density of 'empty' space as a measure of curvature, even if there is an 'ordinary' particle of mass *m* present. Then (3), (7) and (9) yield a simple heuristic relation: $|\Lambda| m^2 = $ constant. This already helps resolve the cosmological-'constant' problem, which is basically the large magnitudes for $\Lambda$ as measures of particle vacuum fields [4]. It can also be further tested, for example by using the Large Hadron Collider.

3. Conclusion

Mach's Principle is broadly understood to mean that the rest mass of a local body should be explained in terms of other material in the universe, rather than being merely given. Einstein was motivated by this Principle in constructing general relativity, but it is widely acknowledged that it cannot be fully realized in such a 4D context. Nevertheless, some recent 4D results on the hierarchy problem are intriguing [1], and have motivated us to re-express the version of 5D noncompactified relativity known as space-time-matter theory [4]. Five-dimensional relativity has been greatly developed in recent years, as a first-order approach to the unification of gravity with the interactions of particles [4]; and it turns out to contain within its formalism a viable account of Machian physics. The two main tricks are to factorize the 4D part of the 5D metric with a quadratic term in the extra coordinate, and to interpret the latter (along with its associated potential) as a measure of particle rest mass. The result, in short, is a momentum manifold rather than a coordinate manifold. It treats mass on the same footing as length and time, and formalizes the law of conservation of energy/momentum.

The interpretation of 5D relativity outlined here depends on shedding the trammels of old Kaluza-Klein theory. The result is an algebraically rich, fully covariant 5D theory. However, to bring the mathematics of this theory into line with known 4D physics requires a choice of gauge, which is a variant of what in the literature is called the canonical metric. In this gauge, it is possible to devise unique tests for the fifth dimension. Four of these have been specified, ranging from particle physics to cosmology.

Acknowledgements

This work grew out of earlier collaborations with B. Mashhoon and J. Ponce de Leon, and has benefited from comments by other members of the STM group. It was partly supported by NSERC.

[19] The fully general equation of motion for a test particle moving in gravitational ($g_{\alpha\beta}$), electromagnetic ($A_\mu$) and scalar ($\Phi$) fields is best stated in terms of a scalar function $n \equiv \varepsilon \Phi^2 (d\ell/dS + A_\alpha dx^\alpha / dS)$. This is a constant of the motion if the 5D metric tensor $g_{AB}$ is independent of $x^4 = \ell$ (ref. 14 pp.160-169). Otherwise, the extra component of the 5D geodesic equation reads $dn/dS = (1/2)(\partial g_{CD}/\partial \ell)(dx^C/dS)(dx^D/dS)$. The four spacetime components of the 5D geodesic read

$$\frac{d^2 x^\mu}{ds^2} + \Gamma^\mu_{\alpha\beta} \frac{dx^\alpha}{ds}\frac{dx^\beta}{ds} = \frac{n}{(1-\varepsilon n^2/\Phi^2)^{1/2}} \left[ F^\mu_{\ \nu} \frac{dx^\nu}{ds} - \frac{A^\mu}{n}\frac{dn}{ds} - g^{\mu\lambda}\frac{\partial A_\lambda}{\partial \ell}\frac{d\ell}{ds} \right]$$

$$+ \frac{\varepsilon n^2}{(1-\varepsilon n^2/\Phi^2)\Phi^3}\left[ \Phi^{;\mu} + \left(\frac{\Phi}{n}\frac{dn}{ds} - \frac{d\Phi}{ds}\right)\frac{dx^\mu}{ds} \right] - g^{\mu\lambda}\frac{\partial g_{\lambda\nu}}{\partial \ell}\frac{dx^\nu}{ds}\frac{d\ell}{ds} .$$

Here the Christoffel symbols account for the 4D curvature, and other notation is the same as in the main text.

[20] The soliton solution is best studied in curvature or Schwarschild-like coordinates (ref. 14 p. 87). The metric is given by $dS^2 = A^a dt^2 - A^{-(a+b)} dr^2 - A^{1-a-b} r^2 (d\theta^2 + \sin^2\theta d\phi^2) - A^b dl^2$. Here $A \equiv (1 - 2M/r)$ defines the source at the centre of the 3D spherically-symmetric space, and the constants $a$, $b$ obey the consistency relation $1 = a^2 + ab + b^2$. The components of the effective energy-momentum tensor $T^\alpha_\beta$ necessary to balance Einstein's equations can be obtained from equation (4) of the main text. In general, the density and pressure fall off asymptotically as $1/r^4$, but these are two special cases which are simpler. For $a = 1, b = 0$ the metric is Schwarzschild-like with an extra flat dimension, and $T^\alpha_\beta = 0$. For $a = 0, b = 1$ the metric has no gravitational effect via its first term but does have an inertial effect via its last term, with $8\pi T^0_0 = 0$, $8\pi T^1_1 = 2M/r^3$, $8\pi T^2_2 = 8\pi T^3_3 = -M/r^3$. For other values of $a$, $b$ the solutions have both gravitational and inertial effects, though in

general not equal. For all *a, b* the trace of $T^\alpha_\beta$ is zero, implying a radiation-like equation of state. For more details, see refs. 14, 15, 16.